# Atom probe analysis of electrode materials for Li-ion batteries: challenges and ways forward

Se-Ho Kim[a], Stoichko Antonov[a], Xuyang Zhou[a], Leigh T. Stephenson[a], Chanwon Jung[a], Ayman A. El-Zoka[a], Daniel K. Schreiber[b], Michele Conroy[c], Baptiste Gault[a,c,*]

- [a.] Max-Planck-Institut für Eisenforschung, Düsseldorf, Germany.
- [b.] Energy and Environment Directorate, Pacific Northwest National Laboratory, P.O. Box 999, Richland, WA 99352, United States
- [c.] Department of Materials, Royal School of Mines, Imperial College London, London, UK

## Abstract

The worldwide developments of electric vehicles, as well as large-scale or grid-scale energy storage to compensate the intermittent nature of renewable energy generation has led to a surge of interest in battery technology. Understanding the factors controlling battery capacity and, critically, their degradation mechanisms to ensure long-term, sustainable and safe operation requires detailed knowledge of their microstructure and chemistry, and their evolution under operating conditions, on the nanoscale. Atom probe tomography (APT) provides compositional mapping of materials in three-dimensions with sub-nanometre resolution, and is poised to play a key role in battery research. However, APT is underpinned by an intense electric-field that can drive lithium migration, and many battery materials are reactive oxides, requiring careful handling and sample transfer. Here, we report on the analysis of both anode and cathode materials, and show that electric-field driven migration can be suppressed by using shielding by embedding powder particles in a metallic matrix or by using a thin conducting surface layer. We demonstrate that for a typical cathode material, cryogenic specimen preparation and transport under ultra-high vacuum leads to major delithiation of the specimen during the analysis. In contrast, transport of specimens through air enables analysis of the material. Finally, we discuss the possible physical underpinnings and discuss ways forward to enable shield of the electric field, which helps address the challenges inherent to the APT analysis of battery materials.

## Introduction

Batteries are at the core of many technologies that will have a significant impact on decarbonation of our society[1]. The high-capacity energy storage needed in electric vehicles or grid energy storage is currently largely achieved using Li-ion batteries (LIBs)[2], which appear one of the most viable and scalable energy storage technology to accommodate the variability of renewable energy sources[3,4], assuming sufficient Li can be extracted. A battery is an assembly of a cathode, an electrolyte and an anode. During charging, the cathode is delithiated and the Li-ions migrate through the electrolyte and are inserted inside the anode; during discharge, the opposite reactions take place[5]. LIBs and their individual constituents have been subject to significant research and development efforts in the past decades[6], leading to the Nobel Prize in Chemistry in 2019 to Goodenough, Whittingham, and Yoshino for their work on Co-based oxides[7,8].

Extending the operation lifetime of LIBs would decrease the environmental footprint and cost, but requires understanding the microstructural origins of capacity-loss and the degradation during cycling to develop strategies to design new high-performance materials[9]. However, these degradation mechanisms occur across length scales ranging from sub-nanometres to microns or more[10,11], and ultimately precluding direct investigation by any single characterization technique. Recent efforts in high-resolution, multiscale, correlative and cryogenic microscopies have been reported, leading to precious insights into these complex processes[12,13]. However, like many oxides, battery materials can be unstable under electron beam irradiation, and transport of samples through air may cause modifications of their composition and structure, making their observation highly challenging[14]. The Li distribution on the nanoscale also remains elusive due to its low atomic weight, which limits its interactions with electrons in transmission electron microscopy (TEM), as well as its high mobility.

Atom probe tomography (APT)[15] has been proposed as a technique to reveal the three-dimensional distribution of Li in these materials. APT provides compositional mapping with sub-nanometre resolution[16,17], and its sensitivity is in principle not related to the atomic weight, hence Li can be readily detected and its distribution analysed. However, there are specific challenges inherent to the analysis of alkali by APT[18–20]. In the case of Li-containing materials, the intense electric field necessary to trigger the field evaporation of surface atoms to enable their analysis can also drive atomic migration during the analysis itself [21]. This *in-situ* delithiation process[22] can make the data unreliable. Species-specific losses associated to the dissociations of molecular ions have been also evidenced in these materials, thereby lowering the spatial resolution and elemental sensitivity[19]. These difficulties have hindered the widespread deployment of the application of APT to materials for LIB and the number of reports has hence remained limited[10,13,23–25].

Here, we report on the atom probe analysis of pristine anode and cathode oxide materials. We show that under conventional laser APT conditions, Li can indeed diffuse out during the analysis. Counter-intuitively, we demonstrate that specimens prepared under cryo-preparation and transferred under ultrahigh-vacuum are particularly subject to this unwanted delithiation, compared to samples prepared and transported under ambient conditions. This outwards migration of Li is driven by the penetration of the electrostatic field inside the specimen[21], we show how the presence of a conducting layer on the specimen's surface provides shielding against the field that can allow to circumvent this problem. In addition, we further show that this layer can be extrinsic, i.e. deposited, or intrinsic as it can develop because of the sensitivity of these materials to an ambient atmosphere. We discuss the details of the origins of the formation of the intrinsic shielding and compare these routes to stability in the APT and the implications for quantification of the data.  Finally, we lay out possible ways forward for the analysis of battery materials by APT.

## Experimental

**Anode Materials**

We first analysed bulk $Li_4Ti_5O_{12}$ (LTO) by atom probe. The microwave-assisted hydrothermal synthesis of these materials is detailed in Ref.[26] and samples were provided by the Dou group at the University of Wollongong, who later demonstrated high performance of these anode materials[27–30]. APT specimens were prepared using a FEI Quanta 3D dual-beam scanning-electron microscope/focused-ion beam (SEM/FIB) and following the in-situ lift-out protocol outlined in Ref.[31]. Specimens were then transferred through air, under ambient temperature and pressure conditions. The data was acquired on a Cameca LEAP 3000X Si at a base temperature of 32K, a laser pulse energy of 0.3 nJ ($\lambda$=532 nm, spot size spot diameter below 10μm), and a detection rate of $2\times10^{-3}$ ion/pulse.

We then analysed a $Li_4Ti_5O_{12}$ commercial powder (>99%, Sigma Aldrich). The powder was first dispersed onto a TEM grid and imaged on the JEM-2200FS TEM (JEOL) operating at 200 kV equipped with an energy-dispersive X-ray analysis (EDX) system (see Figure S1).

The nanoparticle powder was then embedded in a Ni-matrix using the approach described in Ref.[32]. The powder was first dispersed in a Watts Ni-ion electrolyte. Subsequently the solution was poured into a commercial nanoparticle depositor (Oxford Atomic Ltd.). A constant current of -19 mA was applied for 600 s for co-electrodeposition process, which results ~8 μm thickness of the Ni film. The Ni-embedded powder specimens for APT were prepared by a site-specific lift-out standard method on a dual-beam focused ion beam FEI Helios 600 (Thermo-Fisher) followed by ambient air transfer to the atom probe. Details are described in Figure S2. APT measurements were carried out using a CAMECA 5000 XS system (i.e. straight flight path) in pulsed-laser mode at a base temperature of 60 K, a detection rate of 1%, a laser energy of 40 pJ ($\lambda$=355 nm, spot size below 3μm), and a laser frequency of 125 kHz. All measurements had low background level of below 10 ppm/nsec. All APT data were processed and reconstructed with the default voltage-protocol parameters in APSuite developed by CAMECA Instruments.

**Cathode Materials**

Commercial NMC811 was sourced from Targray. Its nominal composition is $LiNi_{0.8}Co_{0.1}Mn_{0.1}O_2$. We used the $N_2$-filled glovebox of the Laplace Project[33] to avoid, or limit the influence of the atmosphere on the specimens following preparation. The powder was dispersed onto Cu tape on a flat-stub for scanning-electron imaging, followed by soft dry-$N_2$ flow to blow off the looser powder particles. The stub was mounted on an atom probe puck sample holder. To avoid contact with air, this puck was then transferred into a dual beam SEM/Xe-plasma FIB (FEI Helios PFIB) through a ultra-high vacuum (UHV)-suitcase. At first, conventional room temperature sharpening of the specimens for APT by was performed. In a second attempt cryogenic sharpening was also employed, as it has been shown to limit damage in beam-sensitive materials[34] and ingress of spurious species[35]. Subsequently, the specimens from both sharpening methods were transferred under UHV, to prevent surface reactions[33], into a CAMECA LEAP 5000 XR, i.e. reflectron-fitted and a CAMECA LEAP 5000 XS. APT analyses were performed across a wide range of conditions of base temperature of 30–60K, laser-pulse energies (5-20 pJ). Trials in high voltage pulsing (5-50%pulse fraction) were also attempted without success.

To mimic results obtained by other groups, specimens from the same material were prepared at room temperature using a Ga ion FIB and transferred through air, under ambient lab conditions. Details are described in Figure S3.

A metallic coating was also deposited directly on specimens using physical-vapor deposition of ultra-high purity nickel (99.99%) mounted in a direct current gun in a BESTEC PVD cluster (BESTEC, Berlin, Germany). The films were grown at a sputtering rate of approximately 0.07nm/s at ambient temperature to a thickness estimated to be 50 nm, directly on the pre-sharpened APT specimens following Ga-FIB preparation and transport through air, under ambient lab conditions. The base pressure prior to sputtering was <6 × 10$^{-6}$ Pa whereupon argon was introduced as the working gas at a flow rate of 20 standard cubic centimetres per minute to a pressure of 0.5 Pa.

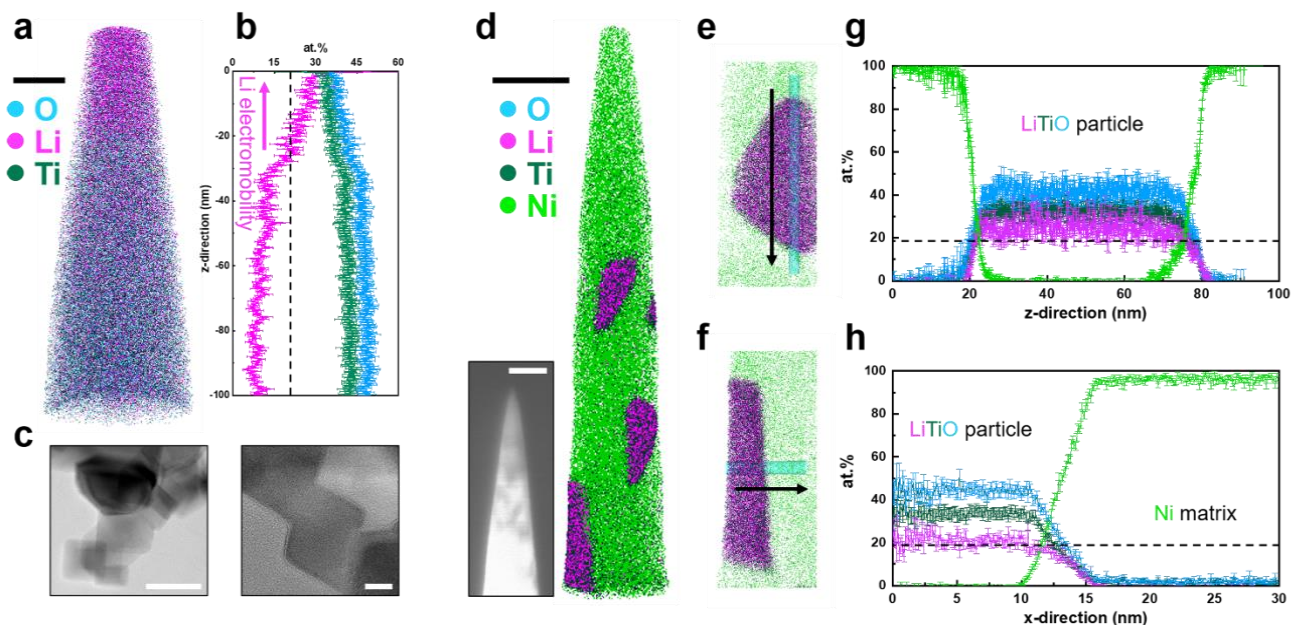

**Figure 1:** (a) 3D atom map of bulk LTO. Cyan, pink, and dark green dots represent reconstructed O, Li, and Ti atoms, respectively. Scale bar is 20 nm. (b) 1D compositional profile along the measurement direction. A black dash-line displays the nominal Li atomic composition. (c) Bright field high resolution TEM images of LTO particles. Scale bars are (left) 50 and (right) 10 nm. (d) 3D atom map of LTO particle embedded in Ni. Scale bar is 50 nm. Green dots represent reconstructed Ni matrix atoms. Inset image is the corresponding backscattered SEM image of LTO particles APT specimen (scale bar = 200 nm). (e)&(f) Extracted nanoparticle regions in the atom map and (g)&(h) corresponding atomic compositional profiles along the cylindrical region of interests (φ5 nm).

All specimens transferred through air, with and without coating, were analysed on a CAMECA LEAP 5000 XS system. From the preceding UHV-cryo transfer APT analyses, the analysis parameters of the moderately successful analyses were a base temperature of 60 K with a pulsed laser energy of 5 pJ (λ=355 nm). Therefore, the same parameters were used for the air-exposed specimens with a pulse frequency of 100 kHz and a detection rate of 0.5% with low background level (<10 ppm/nsec). All the APT data were processed using APSuite and reconstructed using the standard voltage reconstruction protocol[36].

## Results

**Anode materials: bulk LTO**

An example of an APT analysis (LEAP 3000X Si) obtained on the bulk LTO is shown in Figure 1a. A high concentration of Li is readily visible towards the top of the reconstructed data. A one-dimensional composition profile in a 20nm-diameter cylinder parallel to the specimen's main axis is plotted in Figure 1b, showing an increase in the Li content towards the specimen's tip and a constant but relatively low level below 30–40 nm, well below the expected stoichiometry. Multiple datasets collected showed a similar profile, and, based on the experimental and modelling work by Greiwe et al.[21], it was interpreted as evidence for the outwards migration and leaching of Li, i.e. of in-situ delithiation. The specimen's geometry in Greiwe's prior work was rather different, with a thin film of a Li-containing glass deposited on a sharp metallic needle used as a substrate. The high electrostatic field is hence generated at the apex of the metallic substrate, which can facilitate the penetration of the field throughout the film and drive = delithiation. Although this geometry is rather different to our conventional lifted-out specimen, the interpretation proposed by Greiwe et al. seem directly applicable to our observed results of in-situ delithiation of LTO.

**Anode materials: LTO powder**

The LTO powder was first imaged by TEM, as shown in Figure 1c, and the powder appears mostly made of a set of individual cuboidal grains. High-resolution imaging confirms the crystalline nature of the powder grains. Figure S2 is a cross-sectional electron micrograph following FIB slicing of the composite formed by electroplating of the powder within a Ni-metallic matrix to facilitate APT specimen preparation[37]. The dark contrast corresponds to regions containing the LTO powder, and, as shown in the inset image in Figure 1d, some were present inside the specimen following sharpening of the specimen in the FIB at room temperature. Another APT dataset is shown in Figure S4. The Li/Ti atomic ratio in the reconstructed LTO particle is ~0.77, close to the stoichiometry of $Li_4Ti_5O_{12}$ (0.8). The reconstruction from the resulting APT analysis is displayed in Figure 1d, with close-ups on two of the analysed LTO powder grains in Figure 1e and 1f, with morphologies compatible with the TEM imaging. One-dimensional composition profiles obtained from these two grains are plotted in Figure 1g and 1h, respectively, with no sign of delithiation, with a concentration of Li, Ti and O constant across these powder grains.

These results are in stark contrast with those obtained on the bulk LTO, assuming that the results obtained on the LEAP 3000 and 5000 generations are comparable, despite the differences in laser wavelength and spot size. Several studies have demonstrated that data on different generation of instruments[18,38] could be comparable, including for battery materials[19]. Here, the level of background is substantially different (Figure S5), and may well account for some of the Li-losses as discussed in Ref.[18,19]. An important aspect is also that the electrodeposition did not induce changes in the chemical distribution and mass composition of the main species within the powder grains.

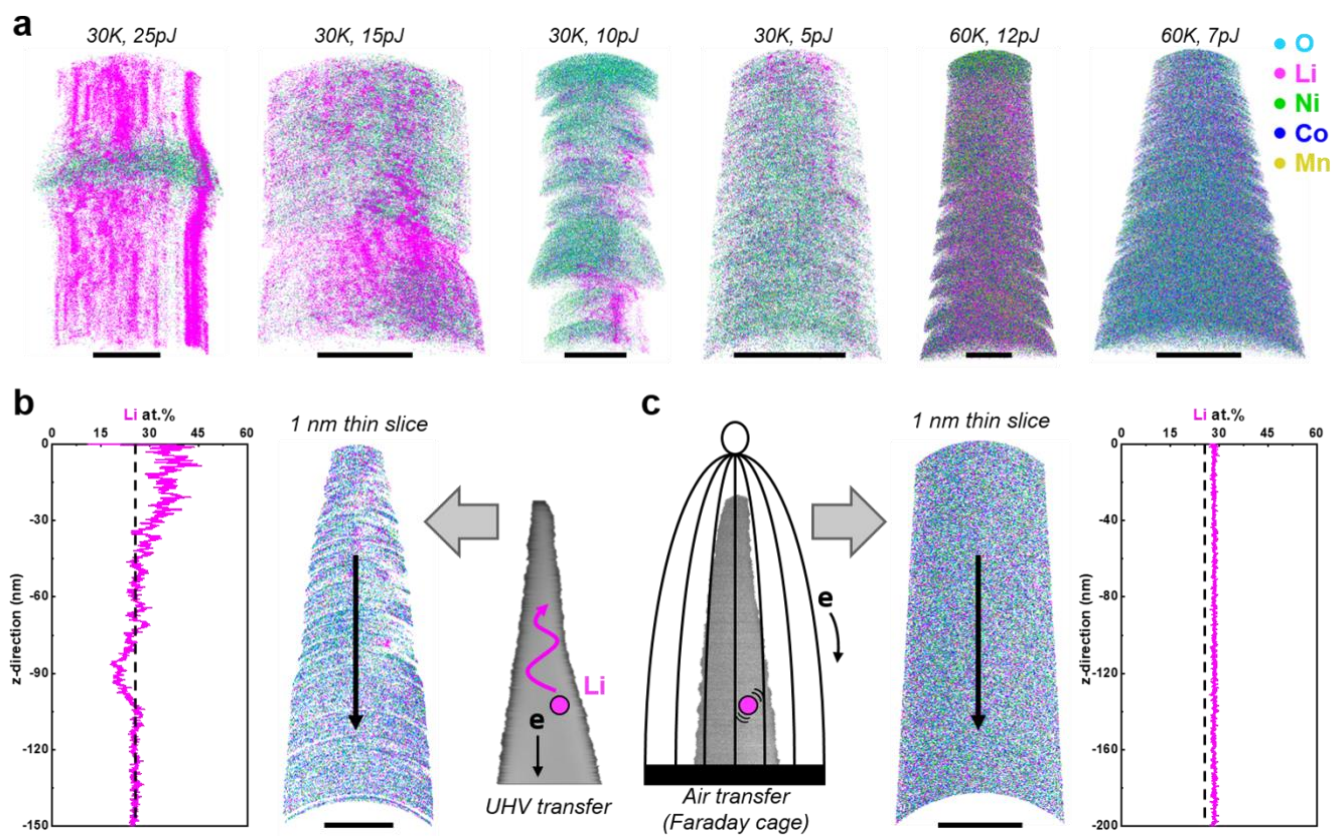

**Figure 2:** (a) A series of 3D atom maps of NMC811 samples after UHV transferring. Scale bars are all 10 nm. A comparison of APT results between (b) UHV- and (c) air transferred samples. The black dash-lines in the 1D composition profiles indicate the nominal Li composition of the NMC811. Both APT measurements were performed at a base temperature at 60 K and a pulsed laser energy of 5 pJ. Scale bars in atom maps are (b) 20 and (c) 50 nm.

**Cathode materials: NMC811 vacuum transfer**

Over two dozen specimens were prepared by lift-out at room-temperature and cryogenic sharpening, followed by ultrahigh-vacuum transfer into the atom probe. In Figure 2a, we display the corresponding tomographic reconstructions from 6 of these datasets, along with the analysis conditions used. All trials led to inhomogeneous field evaporation conditions and regions of (erroneous) high-density of Li indicative of in-situ delithiation. This process also led to unstable field evaporation conditions, as strong bursts of field evaporation of Li led to sudden high detection rates, causing the software to drop the high voltage to lower the electrostatic field and detection rate. These datasets exhibited relatively high levels of background as well (see Figure S5). This appeared to be more pronounced at a lower base temperature (here 30 K) than at higher temperature (e.g. 60 K), i.e. in conditions where the electrostatic field was relatively higher. Dropping the laser pulse energy, that is decreasing the peak temperature reached by the specimen subsequently to the illuminating pulse, also seemed to make the distribution more homogeneous. Nevertheless, the data quality was far from what one could expect or consider acceptable.

Figure 2b displays the results of a single, longer dataset obtained at 60K, 5pJ, and 5 ions per 1000 pulses on average, which appeared to be an optimal set of experimental conditions for that specimen, following the aforementioned preliminary trials. Although the overall bulk composition matches with the nominal composition of the NMC811, even here, the composition profile along the length of the specimen evidences in-situ delithiation, accompanied with strong variations of the point density across the reconstructed dataset.

**Cathode materials: NMC811 atmospheric transfer**

Seeking to reproduce the results from the literature[24,25], we prepared specimens from NMC811 by FIB milling at room temperature and transferred the specimens through air, under ambient lab conditions. Figure 2c displays the corresponding reconstruction from a dataset obtained in laser pulsing mode using the same conditions as for the data reported in Figure 2b. Stable analysis conditions are achieved with 45 x $10^6$ ions collected, the Li distribution appears homogeneous, and there is no evidence for in situ delithiation in the composition profile along the depth of the reconstructed data.

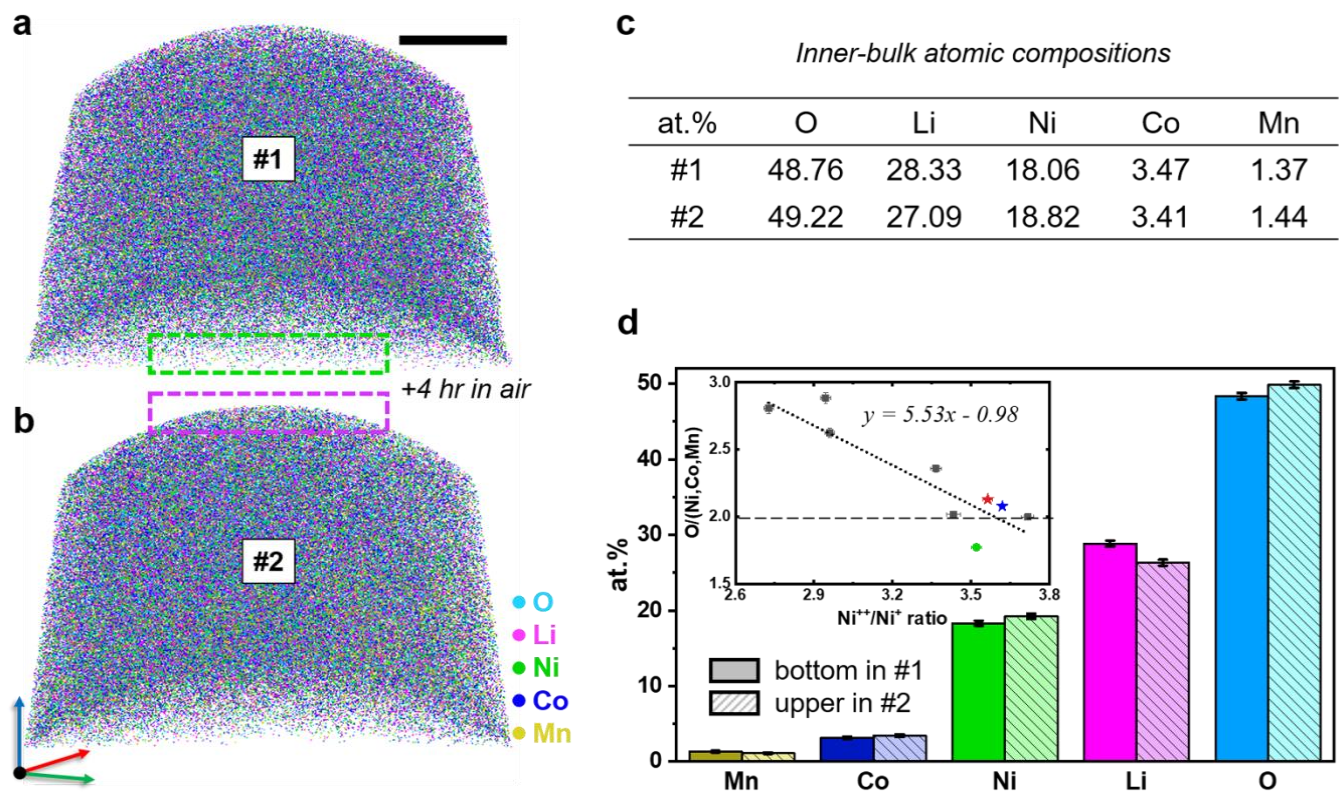

**Figure 3:** 3D atom maps of a sample analysed (a) before and (b) after 4 h oxidation. Scale bar is 20 nm. The measurement was stopped after the first run and the sample was taken out of the high vacuum into air for 4 h. After the oxidation, the measurement was continued without changing the laser position or analysis conditions. 10nm-thin-sliced tomograms of each 3D atom map are presented in Figure S6. (c) Atomic compositions extracted inside each atom map. (d) Composition comparison of atomic layers between before and after oxidation. Inset image shows field strength ($Ni^{++}/Ni^+$) vs. oxygen/metals ratio. The oxygen/metal ratio decreases as the field increases. Red and blue stars are before and after the oxidation, respectively, from 5000 XS measurement and green circle is the data acquired from 5000 HR measurement.

In search of possible compositional changes associated to the air exposure, we performed another similar experiment - reproducing the results in Figure 2c, and the APT reconstruction shown in Figure 3a. We collected over 11 x $10^6$ ions, corresponding to an analysis of 40-50 nm in depth, and observed no substantial changes in the composition of the material during the acquisition. At this stage, the specimen's surface is laid bare and the bulk of the specimen exposed. The specimen was thus taken out of the ultrahigh vacuum of the atom probe, left in ambient lab air for 4 h and reinserted into the APT vacuum chamber. APT data collection from this specimen was then resumed, at a high-voltage comparable to the voltage reached at the end of the first analysis, and a further 13×$10^6$ ions were collected (Figure 3c). The composition of the inner bulk region for each dataset is summarized in Figure 3. The corresponding composition at the end of the first run and the start of the second run are summarised in Figure 3d, suggest a slight enrichment in Ni and O, and decrease in Li at the surface, after exposure to air.

Figure S6 shows 2nm-thick slices through the tomographic reconstructions from the corresponding analyses showing only the Li ion distributions, which appears close to random, and is further confirmed by the close match between the experimental frequency distribution[39] for dataset #2 and the corresponding binomial distribution shown in Figure S7.
However, after the exposure to air of the cleaned NMC811 surface, hydroxyl ions ($OH^+$, 17Da) appear adsorbed on its surface. A third dataset was acquired after this same specimen was taken out of the atom probe ultrahigh vacuum and left in air for 24 h, and again re-inserted and reanalysed. This new experiment started at a voltage similar to the end of the preceding analysis. The specimen only yielded a further 1.5×$10^6$ ions before failing (see Figure S8a and 8b). However, the Li distribution also appears random, unlike results for the UHV-transferred specimens (see Figure S8c), and the absorption of OH-molecular ions on the surface was again observed from air exposure.

Finally, to facilitate the comparison of the composition across datasets, we performed a series of analyses on another specimen, also transported through air, for a range of laser pulse energies, from 2 to 15 pJ at 60K on the same specimen analysed on the LEAP 5000 XS. Indeed, compositional measurements by APT are known to be affected by the strength of the electrostatic field[40,41], even more so for oxides[19,42], through potentially complex loss mechanisms[42–44]. The specific case of the compositional dependence on the intensity of the electrostatic field in mixed NiMnCo-based oxides has not been reported. When the laser energy increased up to 20 pJ, mostly only Li ions were detected, which can be attributed to heat-induced, electric-field-driven delithiation (see Figure S9); therefore, laser energies lower than 15 pJ were used for evaluating the influence of the electric field strength on the composition. We acquired a series of datasets containing approx. $1 \times 10^6$ ions. For each condition, we estimated the charge-state-ratio (CSR) for nickel, defined here as the ratio of the number of ions in the peak of $^{58}Ni^{2+}$ to the number of ions in $^{58}Ni^{1+}$. The CSR is used as a proxy for the intensity of the electrostatic field in the vicinity of the field emitter[45]. The ratio of O to metal (Ni, Mn, Co) is plotted against the CSR in the inset in Figure 3d, along with a linear fit as a guide for the eye. The change in the O concentration is consistent with other reports of the measurement of O composition as a function of the electric field[19,44,46].

## Discussion

### Origins of the in situ delithiation

The in-situ delithiation can be considered as electromigration, i.e. a thermally-activated transport process in which the electric field lowers the activation potential energy barrier, facilitating directional ions migration by thermal agitation. Species with lower energy barriers to movement will therefore migrate under such conditions. It is worth noting that electromigration can drive transport even *against* a concentration gradient – and is typically fast compared to diffusion[47].

Reports of the in-situ delithiation during APT raised questions on its origin:
(i) Is the delithiation taking place even prior to field evaporation during APT data collection and is it associated to the specimen preparation in the SEM/FIB?
The damage and energy imparted by incident electrons during imaging and the ion-beam sputtering / implantation during specimen shaping could indeed have activated the outwards transport of Li towards the surface. This fact motivated the use of the cryogenic specimen preparation methods. In addition, the known sensitivity to air exposure of these reactive materials motivated the use of the vacuum transfer system. This approach, whilst minimising any preparation artefacts, was unsuccessful in mitigating challenges to the analytical performance of APT, albeit only the ultra-high vacuum transfer appears to be to blame.

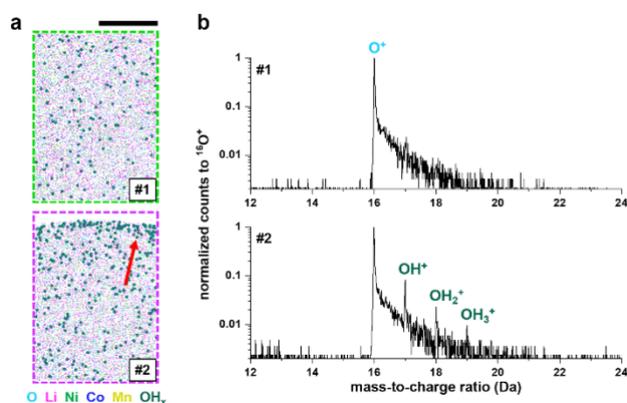

**Figure 4.** (a) $OH_x$ molecules distributions (#1) before and (#2) after the oxidation test on the NMC811 sample. Sizes of the extracted volumes from Figure 3a and 3b are $20 \times 10 \times 30$ nm$^3$. The red arrow indicates the adsorption of $OH_x$ on the 4hr air-exposed surface. (b) The corresponding $^{16}O^+$-normalized mass spectra. Scale bar is 10 nm.

(ii) Is delithiation predominantly, or exclusively, an effect of the electrostatic field that enables ion migration at the low temperature of APT analyses (30–80K typically)[21]?
Our work supports this second mechanism, and demonstrates the importance of shielding the electrostatic field in order to obtain satisfactory results in the analysis of two Li-containing materials that can be subject to in-situ delithiation during APT analysis. Effective shielding can be achieved by coating the specimen with a conductive material for instance. Coating of sharpened specimens had been already demonstrated by several groups, in order primarily to improve yield and the analytical performance of laser-pulsed APT through a faster heat conduction within the coated layer[48–50]. Here, for the LTO, we exploit the shielding offered by an electro-conductive layer to avoid critical issues associated to the penetration of the field within the bulk of the specimen that will drive electro-migration and delithiation during the analysis itself. Exposure of the specimen's surface to a suitable environment may also lead to reactions that modify its surface electronic properties that provide sufficient shielding. This will be material dependent and it may also degrade the properties and subsequent analytical performance. These aspects are further discussed in the following sections.

### Intrinsic shielding

The case of the NMC811 is rather complex and somehow puzzling: the transport of prepared specimens through air appears to lead to an intrinsic shielding that does not form when the specimens are transported under vacuum. There are multiple differences in the two sets of experiments and below, we try to address these successively in the following.

First, the batch of specimens transported in ultra-high vacuum were prepared using a PFIB at cryogenic temperature (Figure 2a–b), whereas the second batch was prepared at room temperature using a conventional Ga-FIB (Figure 2c, Figure 3). To evaluate if the FIB-specimen preparation process was at the origin of the shielding effect, since Ga is known to be chemically active, we performed two new sets of experiments.

Following the sharpening of the specimen at 30kV, we did not perform any cleaning at low acceleration voltage (5kV), and then performed cleaning for 5 seconds and 45 seconds, on three specimens. These specimens were then analysed in the LEAP 5000XS, and the corresponding reconstructed data are shown in Figure S10. We can report that without cleaning, the surface composition contains Ga. However, as the cleaning proceeds, this layers progressively disappears. In all three cases, the specimen was successfully analysed by APT without noticeable delithiation artifacts.

We then prepared specimens with the PFIB at room temperature with ambient air transfer, and Figure S11 is the corresponding reconstructed dataset. Once again, the analysis was successful and the Li is randomly distributed. The difference in behaviour is evidenced by the difference in the voltage curves during the analysis, as shown in Figure S12.

Second, there is a possibility that the composition has changed across the entire specimen. We compared the composition obtained following ultra-high vacuum transfer and air transfer to check for a possible change in composition across the entire specimen. For one of the rare acceptable datasets after UHV transfer, we estimated the O to metal ratio and Ni CSR. This data point was added as a green star in the plot in the inset in Figure 3d. The O content may appear slightly lower, yet as this specimen was analysed on the LEAP 5000 XR, the data may not be directly comparable[51]. The field evaporation conditions appear far from optimal and may lead to additional O losses that cannot be traced easily because of the reflectron that prevents the use of e.g. correlation histograms[52].

The O pick-up could have taken place either during the specimen preparation at room temperature inside the (P)FIB or during transport through air. Hydrogen pick-up was demonstrated in some metallic materials, and was suppressed by using low-temperature[35,53] FIB specimen preparation. This was not reported in oxides though, and the extremely low levels of O-contamination found following cryo-PFIB preparation of APT specimens from pure Mg may dismiss this possibility[33]. The transfer of the specimens through air lasted less than 1h, during which oxygen can adsorb on the surface of reactive metals[33]. An estimate of the time to form a monolayer of $O_2$ is approximately 0.3s in air, compared to nearly 70 days in the UHV conditions used during the transfer (see SI).

We cannot, at this stage conclude whether this could have led to a change in the composition of the material, but these processes would be diffusion controlled and no indication of inward or outward diffusion of species was observed in the elemental distribution extracted from our analyses. In addition, no redistribution of Li or other species was reported following cryogenic specimen preparation and observation by cryo-transmission-electron microscopy[12], this hence seems unlikely.

Another possibility is that only a layer of slightly different composition or structure appears in the near surface region by migration of species from the material itself, possibly activated by the exposure to air (change of the $O_2$ partial pressure). This layer could offer a sufficiently high conductivity to shield the electrostatic field and prevent field penetration. The actual conductivity of such mixed oxides, esp. at low temperature, is often unknown, and it might be that only the first few atomic layers are modified. An increased Ni concentration was shown to increase the conductivity of bulk Ni-Mn-Co oxides for instance[54], even if the resistivity remains orders of magnitude higher than that for a metal. The composition of the surface after the specimen was cleaned by field evaporation and after exposure to air are reported in the inset in Figure 3d as a red and blue star, respectively. The change in the charge-state-ratio from dataset to dataset indicates that the slight difference in the measured composition after exposure to air may be related to the measurement and not to a substantial change in the surface's composition following exposure to air.

Finally, there is a possibility that only the very surface composition was changed by adsorption of oxygen, nitrogen or moisture from the air on the outermost surface of the specimen. The latter could explain the detection of additional $OH_x^+$ at 17-19 Da on the surface after exposure to air (see Figure 4a and 4b). The APT analysis has a limited field of view and cannot collect data from the edges of the specimen[55], hence we can only assume that the entire surface area of the specimen is covered.

The recent efforts around specimen coating using only a few graphene sheets[50,56,57] demonstrates that even very thin conductive layer can achieve the necessary increase in conductivity and sufficient shielding to enable APT analysis. These species may simply neutralise dangling bonds for instance or modify the very local conduction properties, enabling surface conductivity, i.e. akin to a topological insulator. Moisture was shown to cause a bending of the electronic bands and change the surface electronic structure of topological insulators[58]. The bending of the electronic bands can also arise from the intense electrostatic field, which was suggested to enable field-ion imaging of oxides for instance [59–61]. Beyond the NMC811, the notable air-sensitivity of Li-Fe-PO$_4$[62] could also explain the absence of delithiation reported by Santhanagopalan et al.[19].

Measuring the changes in the electronic structure, resistivity or conductivity associated to the slight changes in the surface composition can be extremely challenging. Perhaps X-ray photoelectron spectroscopy or scanning-tunnelling microscopy performed directly on sharpened specimens transferred under UHV conditions and then exposed to air could reveal these changes. Dedicated atomistic simulations may also provide very valuable insights, yet performing accurate e.g. density-functional theory calculations on oxides under the extreme fields encountered during an atom probe analysis is not without challenges. These could constitute important next steps in our study, but fall outside the scope of the current paper. Yet there is evidence from adsorption experiments[63] and density-functional theory[64,65] that gaseous absorption on NiO surfaces makes them more conductive. These appear to support our proposed mechanism, and involve only minute changes to the pristine material underneath the top layer and hence enabling meaningful APT analyses.

**Extrinsic shielding**

We expect, however, processes leading to intrinsic shielding to be very material-dependent and one should hence be careful with interpreting data too hastily. For instance, the compositional or structural modification of the material upon cycling may prevent the formation of this conductive layer, thereby preventing the study of degradation processes or of slight changes in composition of the surface that could explain the drop in lifetime. For instance, despite transport through air, the results on the LTO contrast with what we observed in the case of the reactive NMC811. We could interpret these results as LTO being inert in air, and as such requiring an extrinsic conductive coating. There is also a possibility that the artefacts observed were due to the use of an older generation of instruments that was notable for facing problems in detecting alkali[18].

Ultimately, LTO is typically used for the anode, and we were able to simply use the electroplating approach, which has demonstrated its versability for the analysis of a range of nanostructures[66–68]. Electroplating could however only be used in cases where the application of the voltage will not affect the atomic distribution or its composition, as some material can be subject to corrosion or dealloying from the combined effect of the solution and applied potential. Extrinsic shielding through the use of a capping layer directly deposited on the specimen would be more reliable for routine analysis.

We hence tried to deposit metal by PVD on already sharpened specimens to demonstrate the feasibility of protecting the surface yet achieving appropriate analysis conditions, as had been demonstrated previously by Seol et al.[48] for instance. The PVD was performed following transport through air though, and led to satisfactory analysis, as shown in Figure S13 and S14. The measured composition of the surface and bulk, and well as the Li distribution appear comparable to the data reported in Figure 2c and Figure 3. Any modification of the material's surface or composition has already taken place.

Going back to the discussion point above of a possible change in the very surface composition via adsorption, and that the shielding can be achieved extrinsically also for NMC811, we would need to find a way to perform metal deposition reachable through UHV-transport or directly inside the FIB. The chemical-vapour deposition from the gas-injection-system inside the FIB may also be an option for coating specimens[69], yet the field evaporation behaviour of the deposited precursor rarely enables satisfactory analysis conditions[70]. In addition, FIB-based deposition under cryogenic conditions requires much fine-tuning to be achieved properly[71,72], and may need to be followed by an appropriate level of in-situ curing via illumination by the electron and/or ion beam[73]. We show, however, that specimens can yield APT data following PVD and that the PVD does not lead to a modification of the specimen's composition, which in principle, validates this approach for extrinsic shielding.

## Conclusions

To summarise, we have shown that the analysis of Li-containing materials by APT is extremely challenging due to the influence of the electrostatic field applied during the experiment that drives in-situ delithiation. The variations of the experimental conditions do not lead to any satisfactory solution. We also show that the use of cryogenic preparation and transfer are not always the "miracle solution" that we sometimes hope it will be for the analysis of air-sensitive materials. We have however demonstrated that by shielding the field through building the equivalent of a Faraday cage at the specimen's surface, we can prevent penetration of the field and hence prevent delithiation. Whether or not a coating is necessary will be highly dependent on the nature of the considered specimen, as the surface reactivity of the NMC811 led to the formation of an intrinsic shielding, or on the goal of the analysis, as the intrinsic shielding alters the surface of the sample. How much this process may have been taking place without being noticed and enabled previous APT analyses of oxides could also be debated at length in the future. Generalising the use of an extrinsic shielding by either metal electrodeposition or physical-vapor deposition could facilitate the APT analysis of some of these highly challenging materials, maybe also through future workflow to combine with cryogenic preparation that will need to be developed.

## Conflicts of interest

There are no conflicts to declare.

## Acknowledgements

Drs Wai-Kong Yeoh and Xun(Joe) Xu are gratefully acknowledged for allowing us to use the data they obtained as part of their preliminary study on LTO. This data was collected in 2011 at the University of Sydney, as part of a trial which was then stopped because of the reliability of the data following in-situ delithiation. We therefore acknowledge technical support from the AMMRF (ammrf.org.au) node at the University of Sydney for the work performed at the time. Daisy Thornton, Dr. Ifan Stephens and Prof. Mary P. Ryan are gratefully acknowledged for fruitful discussions and provision of the NMC811 samples. We thank Uwe Tezins, Christian Broß and Andreas Sturm for their support to the FIB and APT facilities at MPIE. We are grateful for the financial support from the BMBF via the project UGSLIT and the Max-Planck Gesellschaft via the Laplace project. S.-H.K., A.A.E., L.T.S. and B.G. acknowledge financial support from the ERC-CoG-SHINE-771602. S.A. and X.Z. are grateful for funding from the AvH foundation. Dr James Best (MPIE) is gratefully acknowledged for his support with the PVD. D.K.S. acknowledges support from the U.S. Department of Energy (DOE) Office of Science, Basic Energy Sciences, Materials Science and Engineering Division. Pacific Northwest National Laboratory is a multiprogram laboratory operated by Battelle for the U.S. DOE under Contract DE-AC05-79RL01830. S.C. acknowledges support from the Royal Society through the University Research Fellowship scheme.

## References

1  J. M. S. L. P Albertus, *Joule*, 2020, **4**, 21–32.
2  M. A. JM Tarascon, *Nature*, 2001, **414**, 359–367.
3  J. T. M Armand, *Nature*, 2008, **451**, 652–657.
4  R. M. R. E. G. S. D. A. V Etacheri, *Energy Environ. Sci.*, 2011, **4**, 3243–3262.
5  Y. M. J. B. C. G. G. C. K Kang, *Science (80-. ).*, 2006, **311**, 977.
6  A. Manthiram, *ACS Cent. Sci.*, 2017, **3**, 1063–1069.
7  K.-S. P. JB Goodenough, *J. Am. Chem. Soc.*, 2013, **135**, 1167–1176.
8  P. J. P. W. J. G. K Mizushima, *Mater. Res. Bull.*, 1980, **15**, 783–798.
9  N. Y. Kim, T. Yim, J. H. Song, J. S. Yu and Z. Lee, *J. Power Sources*, 2016, **307**, 641–648.
10 D. Mohanty, B. Mazumder, A. Devaraj, A. S. Sefat, A. Huq, L. A. David, E. A. Payzant, J. Li, D. L. Wood and C. Daniel, *Nano Energy*, 2017, **36**, 76–84.
11 S.-K. Jung, H. Gwon, J. Hong, K.-Y. Park, D.-H. Seo, H. Kim, J. Hyun, W. Yang and K. Kang, *Adv. Energy Mater.*, 2014, **4**, 1300787.
12 M. J. Zachman, Z. Tu, S. Choudhury, L. A. Archer and L. F. Kourkoutis, *Nature*, 2018, **560**, 345–349.
13 B.-G. Chae, S. Y. Park, J. H. Song, E. Lee and W. S. Jeon, *Nat Commun.*, 2021, **12**, 3814.
14 L. Yu, M. Li, J. Wen, K. Amine and J. Lu, *Mater. Chem. Front.*, 2021, **5**, 5186–5193.
15 B. et al Gault, A. Chiaramonti, O. Cojocaru-Mirédin, P. Stender, R. Dubosq, C. Freysoldt, S. K. Makineni, T. Li, M. Moody and J. M. Cairney, *Nat. Rev. Methods Prim.*, 2021, 1–51.
16 F. De Geuser and B. Gault, *Acta Mater.*, 2020, **188**, 406–415.
17 B. M. Jenkins, F. Danoix, M. Gouné, P. A. J. Bagot, Z. Peng, M. P. Moody and B. Gault, *Microsc. Microanal.*, 2020, **26**, 247–257.
18 X. Lu, D. K. Schreiber, J. J. Neeway, J. V. Ryan and J. Du, *J. Am. Ceram. Soc.*, 2017, **100**, 4801–4815.
19 D. Santhanagopalan, D. K. Schreiber, D. E. Perea, R. L. Martens, Y. Janssen, P. Khalifah and Y. S. Meng, *Ultramicroscopy*, 2015, **148**, 57–66.
20 S. Gin, J. V Ryan, D. K. Schreiber, J. Neeway and M. Cabié, *Chem. Geol.*, 2013, **349–350**, 99–109.
21 G.-H. Greiwe, Z. Balogh, G. Schmitz and Z. B. G. S. G-H Greiwe, *Ultramicroscopy*, 2014, **141**, 51–55.
22 B. Pfeiffer, J. Maier, J. Arlt and C. Nowak, *Microsc. Microanal.*, 2017, **23**, 314–320.
23 J. Maier, B. Pfeiffer, C. A. Volkert and C. Nowak, *Energy Technol.*, 2016, **4**, 1565–1574.
24 A. Devaraj, M. Gu, R. Colby, P. Yan, C. M. Wang, J. M. Zheng, J. Xiao, A. Genc, J. G. Zhang, I. Belharouak, D. Wang, K. Amine and S. Thevuthasan, *Nat. Commun.*, , DOI:10.1038/ncomms9014.
25 J. Y. Lee, J. Y. Kim, H. I. Cho, C. H. Lee, H. S. Kim, S. U. Lee, T. J. Prosa, D. J. Larson, T. H. Yu and J. P. Ahn, *J. Power Sources*, 2018, **379**, 160–166.
26 S.-L. Chou, J.-Z. Wang, H.-K. Liu and S.-X. Dou, *J. Phys. Chem. C*, 2011, **115**, 16220–16227.

27 J.-G. Kim, D. Shi, M.-S. Park, G. Jeong, Y.-U. Heo, M. Seo, Y.-J. Kim, J. H. Kim and S. X. Dou, *Nano Res. 2013 65*, 2013, **6**, 365–372.
28 J.-G. Kim, D. Shi, K.-J. Kong, Y.-U. Heo, J. H. Kim, M. R. Jo, Y. C. Lee, Y.-M. Kang and S. X. Dou, *ACS Appl. Mater. Interfaces*, 2013, **5**, 691–696.
29 D. Su, S. Dou and G. Wang, *Chem. Mater.*, 2015, **27**, 6022–6029.
30 D. C. Hofmann, J. Y. Suh, A. Wiest, G. Duan, M. L. Lind, M. D. Demetriou and W. L. Johnson, *Nature*, 2008, **451**, 1085--U3.
31 P. J. Felfer, T. Alam, S. P. Ringer and J. M. Cairney, *Microsc. Res. Tech.*, 2012, **75**, 484–491.
32 S.-H. S. H. Kim, J. Lim, R. Sahu, O. Kasian, L. T. L. T. Stephenson, C. Scheu and B. Gault, *Adv. Mater.*, , DOI:10.1002/adma.201907235.
33 L. T. Stephenson, A. Szczepaniak, I. Mouton, K. A. K. Rusitzka, A. J. Breen, U. Tezins, A. Sturm, D. Vogel, Y. Chang, P. Kontis, A. Rosenthal, J. D. Shepard, U. Maier, T. F. Kelly, D. Raabe and B. Gault, *PLoS One*, 2018, **13**, e0209211.
34 N. A. Rivas, A. Babayigit, B. Conings, T. Schwarz, A. Sturm, A. G. Manjón, O. Cojocaru-Mirédin, B. Gault and F. U. Renner, *PLoS One*, , DOI:10.1371/journal.pone.0227920.
35 Y. Chang, W. Lu, J. Guénolé, L. T. Stephenson, A. Szczpaniak, P. Kontis, A. K. Ackerman, F. Dear, I. Mouton, X. Zhong, D. Raabe, B. Gault, S. Zhang, D. Dye, C. H. Liebscher, D. Ponge, S. Korte-Kerze, D. Raabe and B. Gault, *Nat. Commun.*, 2019, **10**, 942.
36 D. J. Larson, T. J. Prosa, R. M. Ulfig, B. P. Geiser and T. F. Kelly, *New York, US Springer Sci.*, 2013, 318.
37 S.-H. H. Kim, P. W. Kang, O. O. Park, J.-B. B. Seol, J.-P. P. Ahn, J. Y. Lee and P.-P. P. Choi, *Ultramicroscopy*, 2018, **190**, 30–38.
38 Z. Peng, P.-P. P.-P. P.-P. Choi, B. Gault and D. Raabe, *Microsc. Microanal.*, 2017, **23**, 1–12.
39 M. P. Moody, L. T. Stephenson, A. V Ceguerra and S. P. Ringer, *Microsc. Res. Tech.*, 2008, **71**, 542–550.
40 M. K. Miller, *J. Vac. Sci. Technol.*, 1981, **19**, 57.
41 F. Tang, B. Gault, S. P. P. S. P. Ringer and J. M. J. M. M. Cairney, *Ultramicroscopy*, 2010, **110**, 836–843.
42 B. Gault, D. W. Saxey, M. W. Ashton, S. B. Sinnott, A. N. Chiaramonti, M. P. Moody and D. K. Schreiber, *New J. Phys.*, 2016, **18**, 33031.
43 I. Blum, L. Rigutti, F. Vurpillot, A. Vella, A. Gaillard and B. Deconihout, *J. Phys. Chem. A*, 2016, **120**, 3654–3662.
44 D. Zanuttini, I. Blum, L. Rigutti, F. Vurpillot, J. Douady, E. Jacquet, P.-M. Anglade and B. Gervais, *Phys. Rev. A*, 2017, **95**, 61401.
45 D. R. Kingham, *Surf. Sci.*, 1982, **116**, 273–301.
46 L. Mancini, N. Amirifar, D. Shinde, I. Blum, M. Gilbert, A. Vella, F. Vurpillot, W. Lefebvre, R. Lardé, E. Talbot, P. Pareige, X. Portier, A. Ziani, C. Davesnne, C. Durand, J. Eymery, R. Butté, J.-F. Carlin, N. Grandjean and L. Rigutti, *J. Phys. Chem. C*, 2014, **118**, 24136–24151.
47 T. T. Tsong, *Atom-Probe Field Ion Microscopy: Field Ion Emission, and Surfaces and Interfaces at Atomic Resolution*, Cambridge University Press, Cambridge, 1990.
48 J. B. Seol, C. M. Kwak, Y. T. Kim and C. G. Park, *Appl. Surf. Sci.*, 2016, **368**, 368–377.
49 D. Larson, T. Prosa, J. Bunton, D. Olson, D. Lawrence, E. Oltman, S. Strennin and T. Kelly, *Microsc Microanal*, 2013, **19**, 994–995.
50 F. Exertier, J. Wang, J. Fu and R. K. W. Marceau, *Microsc. Microanal.*, 2021, 1–12.
51 Y. H. Chang, I. Mouton, L. Stephenson, M. Ashton, G. K. Zhang, A. Szczpaniak, W. J. Lu, D. Ponge, D. Raabe and B. Gault, *New J. Phys.*, 2019, **21**, 053025.
52 D. W. Saxey, *Ultramicroscopy*, 2011, **111**, 473–479.
53 I. Mouton, Y. Chang, P. Chakraborty, S. Wang, L. T. Stephenson, T. Ben Britton and B. Gault, *Materialia*, 2021, **15**, 101006.
54 T. A. Ta, H. S. Nguyen, O. T. T. Nguyen, C. T. Dang, L. A. Hoang and L. D. Pham, *Mater. Res. Express*, 2019, **6**, 065505.
55 M. Herbig, *Scr. Mater.*, 2018, **148**, 98–105.
56 V. R. Adineh, C. Zheng, Q. Zhang, R. K. W. Marceau, B. Liu, Y. Chen, K. J. Si, M. Weyland, T. Velkov, W. Cheng, J. Li and J. Fu, *Adv. Funct. Mater.*, 2018, **28**, 1801439.
57 S. Qiu, V. Garg, S. Zhang, Y. Chen, J. Li, A. Taylor, R. K. W. Marceau and J. Fu, *Ultramicroscopy*, 2020, 113036.
58 H. M. Benia, C. Lin, K. Kern and C. R. Ast, *Phys. Rev. Lett.*, 2011, **107**, 177602.
59 G. Liu, G. Zhou and Y.-H. Chen, *Appl. Phys. Lett.*, 2012, **101**, 223109.
60 J. L. Collins, A. Tadich, W. Wu, L. C. Gomes, J. N. B. Rodrigues, C. Liu, J. Hellerstedt, H. Ryu, S. Tang, S.-K. Mo, S. Adam, S. A. Yang, M. S. Fuhrer and M. T. Edmonds, *Nat. 2018 5647736*, 2018, **564**, 390–394.
61 E. P. Silaeva, L. Arnoldi, M. L. Karahka, B. Deconihout, A. Menand, H. J. Kreuzer and A. Vella, *Nano Lett.*, 2014, **14**, 6066–6072.
62 K. Zaghib, M. Dontigny, P. Charest, J. F. Labrecque, A. Guerfi, A. Mauger, F. Gendron and C. M. Julien, *J. Power Sources*, 2008, **185**, 698–710.
63 Y. Inoue, K. Sato and S. Suzuki, *J. Phys. Chem.*, 1985, **89**, 2827–2831.
64 B. Wang, J. Nisar and R. Ahuja, *ACS Appl. Mater. Interfaces*, 2012, **4**, 5691–5697.
65 A. Mohsenzadeh, K. Bolton and T. Richards, *Surf. Sci.*, 2014, **627**, 1–10.
66 J. Lim, S.-H. S. H. Kim, R. Aymerich Armengol, O. Kasian, P.-P. P. P. Choi, L. T. L. T. Stephenson, B. Gault and C. Scheu, *Angew. Chemie - Int. Ed.*, , DOI:10.1002/anie.201915709.
67 K. Jang, S.-H. Kim, H. Jun, C. Jung, J. Yu, S. Lee and P.-P. Choi, *Nat. Commun.*
68 S. H. Kim, K. Jang, P. W. Kang, J. P. Ahn, J. B. Seol, C. M. Kwak, C. Hatzoglou, F. Vurpillot and P. P. Choi, *J. Alloys Compd.*, 2020, **831**, 154721.

69  D. R. Diercks, B. P. Gorman and J. J. L. Mulders, *Microsc. Microanal.*, 2017, **23**, 321–328.
70  S. Gerstl, A. Morrone and R. Kvitek, *Microsc. Microanal.*, 2006, **12**, 1252–1253.
71  D. E. Perea, S. S. A. Gerstl, J. Chin, B. Hirschi and J. E. Evans, *Adv. Struct. Chem. imaging*, 2017, **3**, 12.
72  A. Salvador-Porroche, S. Sangiao, P. Philipp, P. Cea and J. M. De Teresa, *Nanomater. 2020, Vol. 10, Page 1906*, 2020, **10**, 1906.
73  C. D. PARMENTER and Z. A. NIZAMUDEEN, *J. Microsc.*, 2020, jmi.12953.

# Atom probe analysis of electrode materials for Li-ion batteries: challenges and ways forward

Se-Ho Kim[a], Stoichko Antonov[a], Xuyang Zhou[a], Leigh T. Stephenson[a], Chanwon Jung[a], Ayman A. El-Zoka[a], Daniel K. Schreiber[b], Michele Conroy[c], Baptiste Gault[a,c,*]

[a]Max-Planck-Institut für Eisenforschung, Düsseldorf, Germany.

[b]Energy and Environment Directorate, Pacific Northwest National Laboratory, P.O. Box 999, Richland, WA 99352, United States

[c]Department of Materials, Royal School of Mines, Imperial College London, London, UK

[*] Corr. Author: b.gault@mpie.de

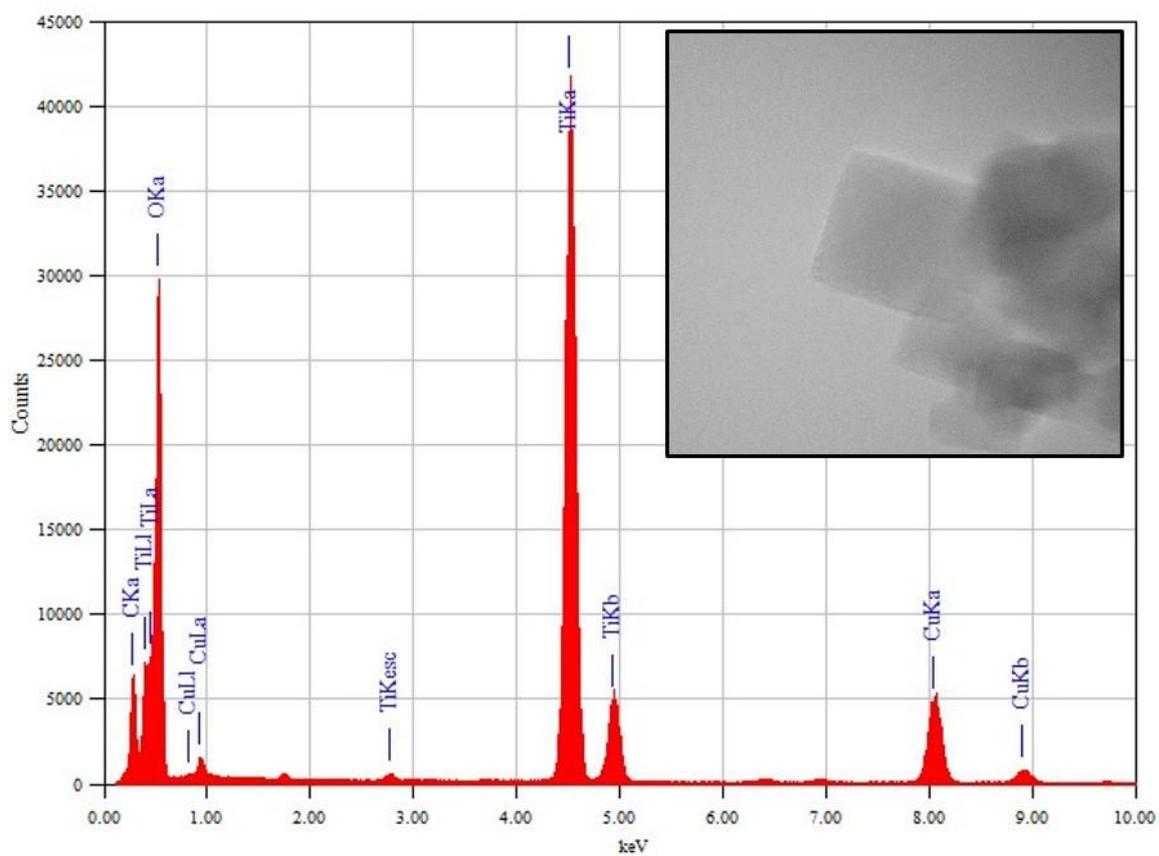

**Figure S1.** EDX spectrum of as-received Li$_4$Ti$_5$O$_{12}$ nanoparticles. Inset image shows the corresponding nanoparticles. Cu and C peaks are originated from a commercial TEM grid. No Li peak detected due to high background signal at low keV range.

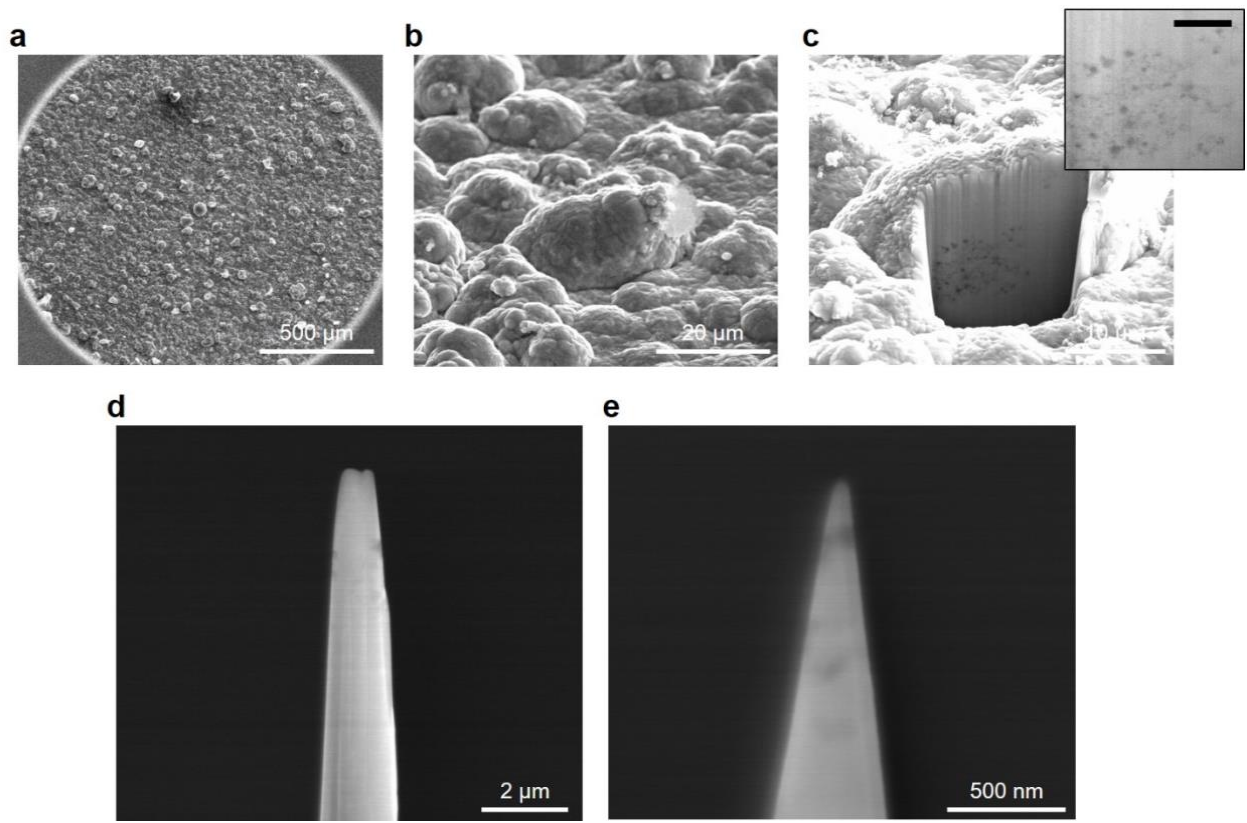

**Figure S2.** APT specimen preparation from the electrodeposition sample. (a) Surface and (b) side-view images of Ni-electrodeposited $Li_4Ti_5O_{12}$ particles from FIB-SEM. (c) Cross-sectional image after Ga-ion beam milling (30 kV, 21 nA). Inset image shows the embedded particles (scale bar = 2 µm). (d) After the standard lift-out, the sample was mounted on a Si micro-post with Pt/C (30 kV, 89 pA). For the annular milling process, an ion-beam voltage of 30 kV and a current of 0.28 nA were used. (d) A final APT specimen was cleaned by low-ion-beam power (5 kV and 8 pA) for 30 sec.

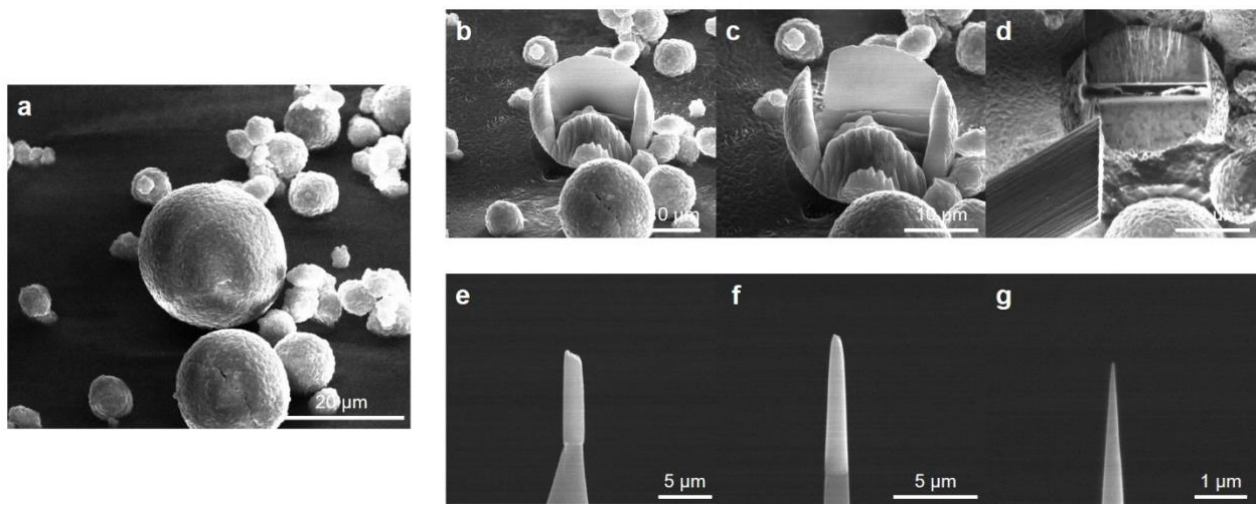

**Figure S3.** APT specimen preparation from NMC811 particles. (a) SEM-FIB image of as-received NMC811 particles. (b) Trenches from front and back-side Ga millings (30 kV, 21 nA) (c) L-shape cut on the left and bottom side from the lamella. (d) Lift-out process using a micro-omniprobe. (e) Mounted NMC811 sample on a Si post. (f) Annular milling at 30 kV and 0.28 nA. (g) Cleaning process (5kV, 8pA) to get rid of highly damaged regions and/or possible Ga implementation.

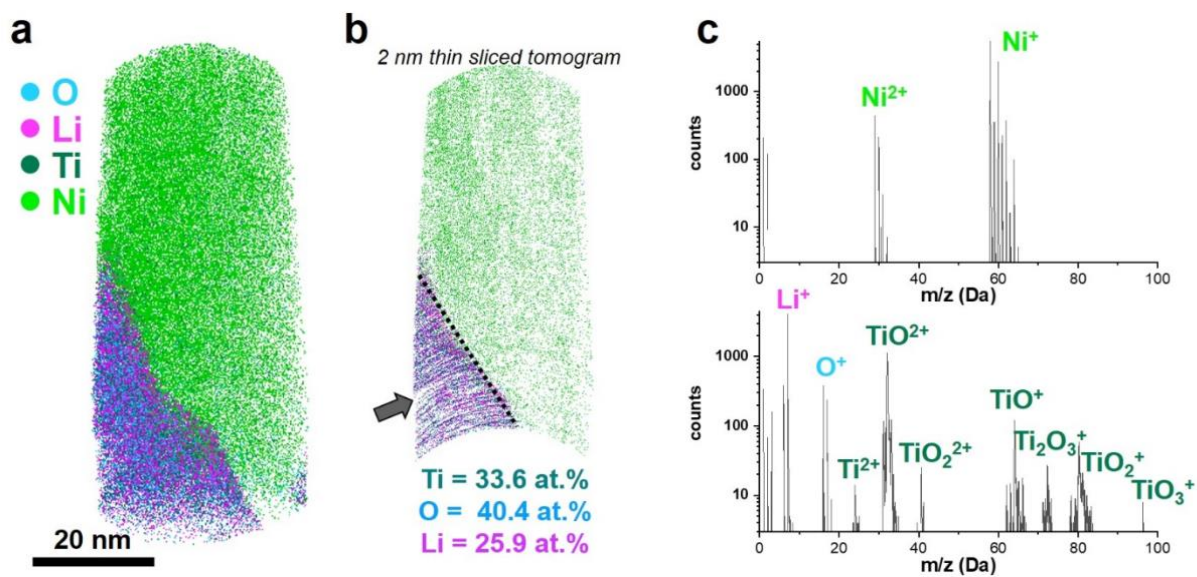

**Figure S4.** (a) 3D atom map of $Li_4Ti_5O_{12}$ particles in Ni matrix. (b) Tomogram from Figure S3a. Gray arrow indicates that there was micro-facture during the APT measurement. (c) Mass spectra from (top) Ni matrix and (bottom) particle region. Cyan, pink, dark green, and green dots represent reconstructed O, Li, Ti, and Ni atoms, respectively.

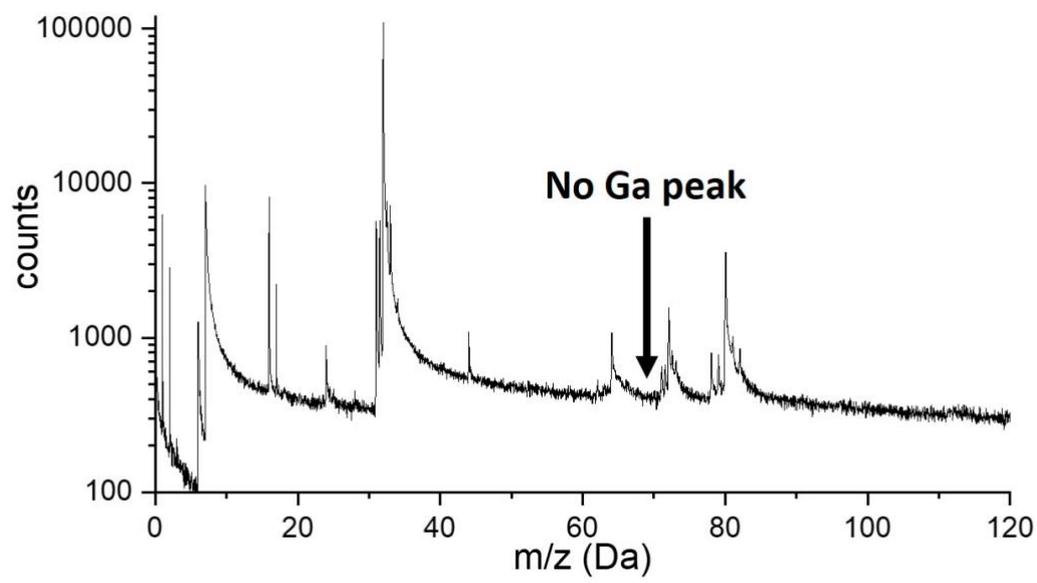

**Figure S5.** Mass spectrum of LTO bulk acquired from 3000 series atom probe. Note that no Ga peak at 69 Da detected.

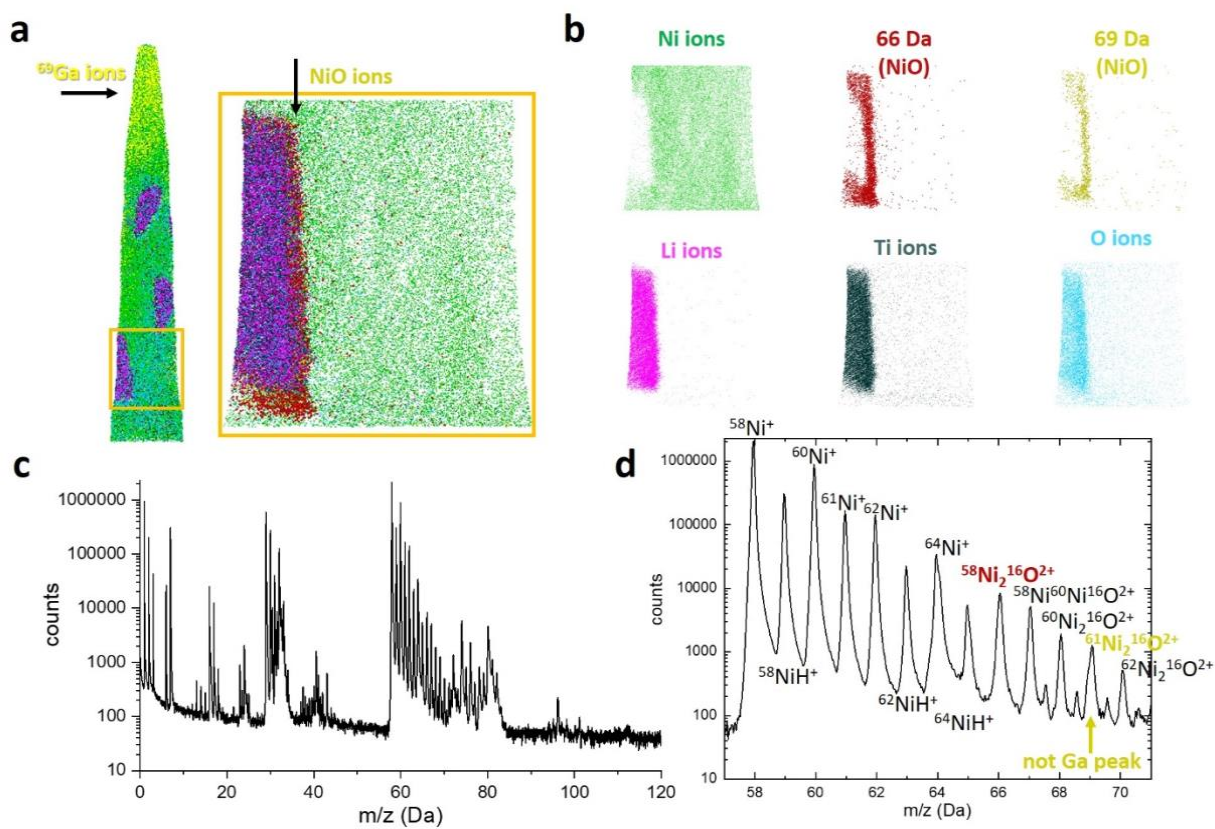

**Figure S6.** (a) 3D atom map and extracted ROI of LTO-Ni composite. Note that uncleaned Ga was detected at the beginning of the measurement. (b) Atom map of each ion: Ni, Li, Ti, O, 66 Da and 69 Da ions from the orange ROI. (c) Mass spectrum of extracted region. (d) Selected mass spectrum region between 55-71 Da. The peak at 69 Da was detected first presumed Ga ions however the consecutive peaks at 66, 67, 68, 70 Da and minor peaks at 67.5, 68.5, 69.5 Da implies the 69 Da most likely corresponds to a minor peak of the NiO layer that was developed during the electroplating (Kim et al. Ultramicroscopy 190 30-38).

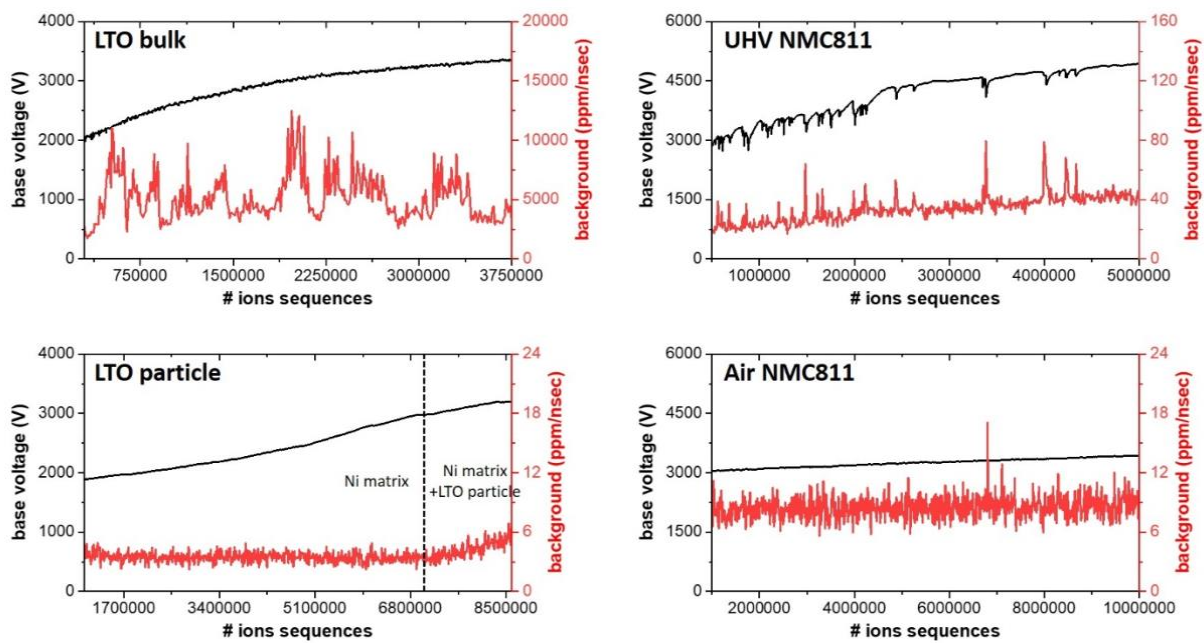

**Figure S7.** Voltage (black) and noise level (red) curves *vs.* ions sequences on each representative APT measurement of the Li battery materials. Note that the LTO bulk was measured with green-laser assisted 3000 HR. In the LTO particle plot, only Ni matrix ions were detected for first 7M ions until the LTO embedded particle appeared.

For the LTO, since the majority of the data acquired is form the metallic matrix, we also ranged an arbitrary peak at 5 Da (no peak appeared) for both bulk and particle (extracted) LTO datasets at same sampled-ion counts. The LTO bulk dataset had 0.1427 at.% background concentration and for LTO particle, it showed 0.038 at.%. For another an arbitrary peak at 150 Da, the LTO bulk shows 0.6086 at.% and the LTO particle: 0.007 at.%. Overall, this indicates a much higher level of background in the LTO data from the bulk acquired on the LEAP 3000X Si, compared to LEAP 5000 XS.

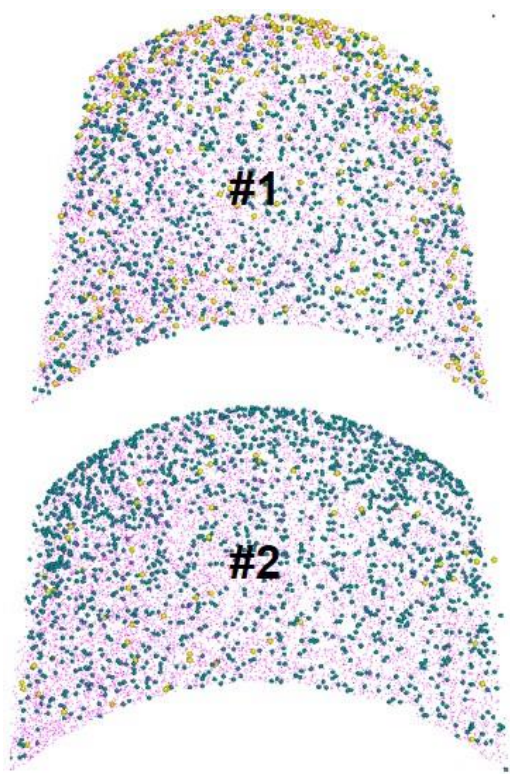

Figure S8. OH$_x$ ions (blue-green) distributions in 10nm-thin sliced atom map from (up) Figure 3a and (down) Figure 3b. Yellow and pink dots represent the reconstructed Ga and Li atoms, respectively.

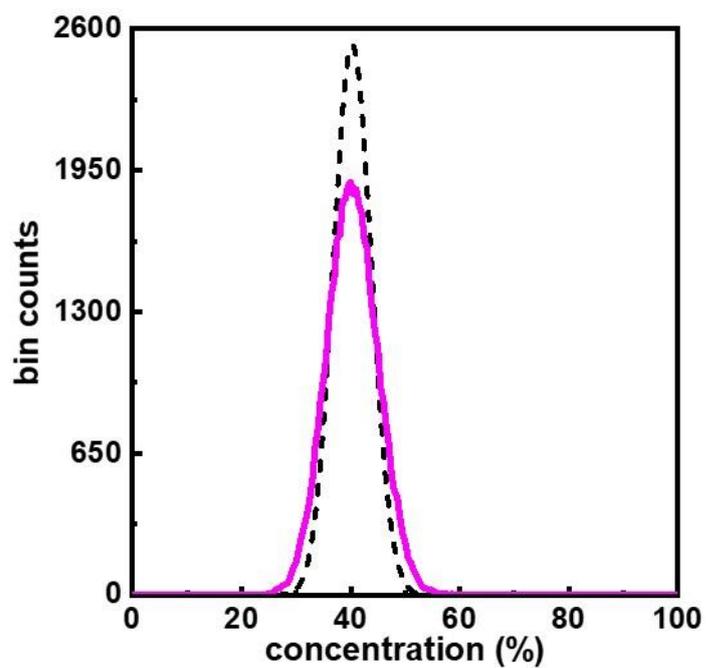

**Figure S9.** (f) Measured binomial distribution of Li in ion% (pink line) compared to an expected random distribution (dashed line).

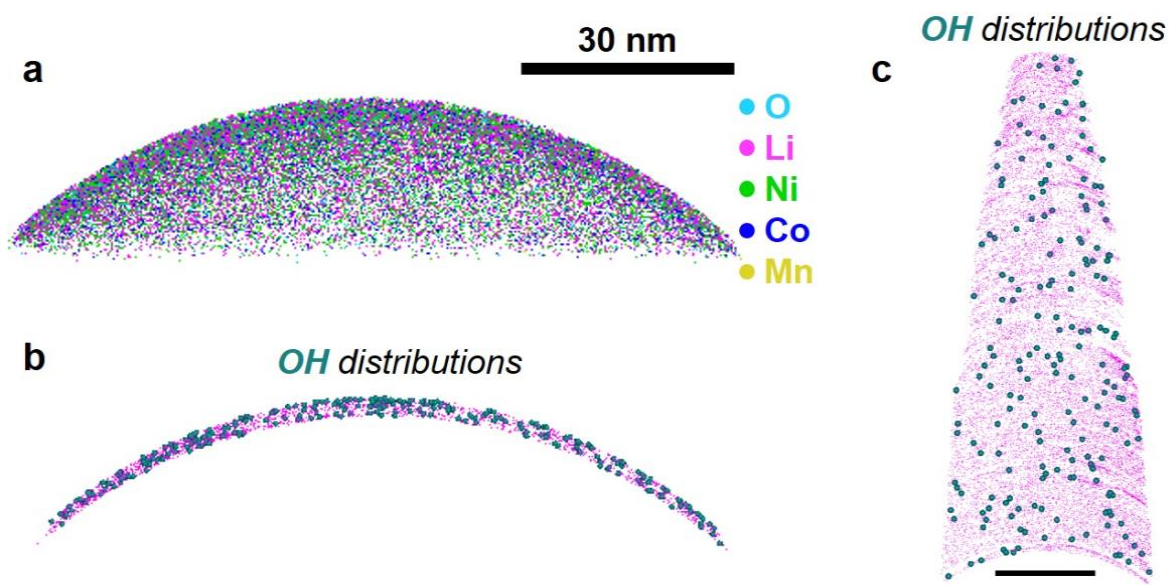

**Figure S10.** The NMC811 APT result after 24 hr oxidation experiment: (a) 3D atom map and (b) 10nm thin sliced tomogram showing detected OH molecular ions distributions. Cyan, pink, green, blue, and yellow dots represent reconstructed O, Li, Ni, Co, and Mn atoms, respectively. (c) 10nm thin sliced tomogram from the UHV-transferred NMC sample (Figure 2b) showing OH ions distribution.

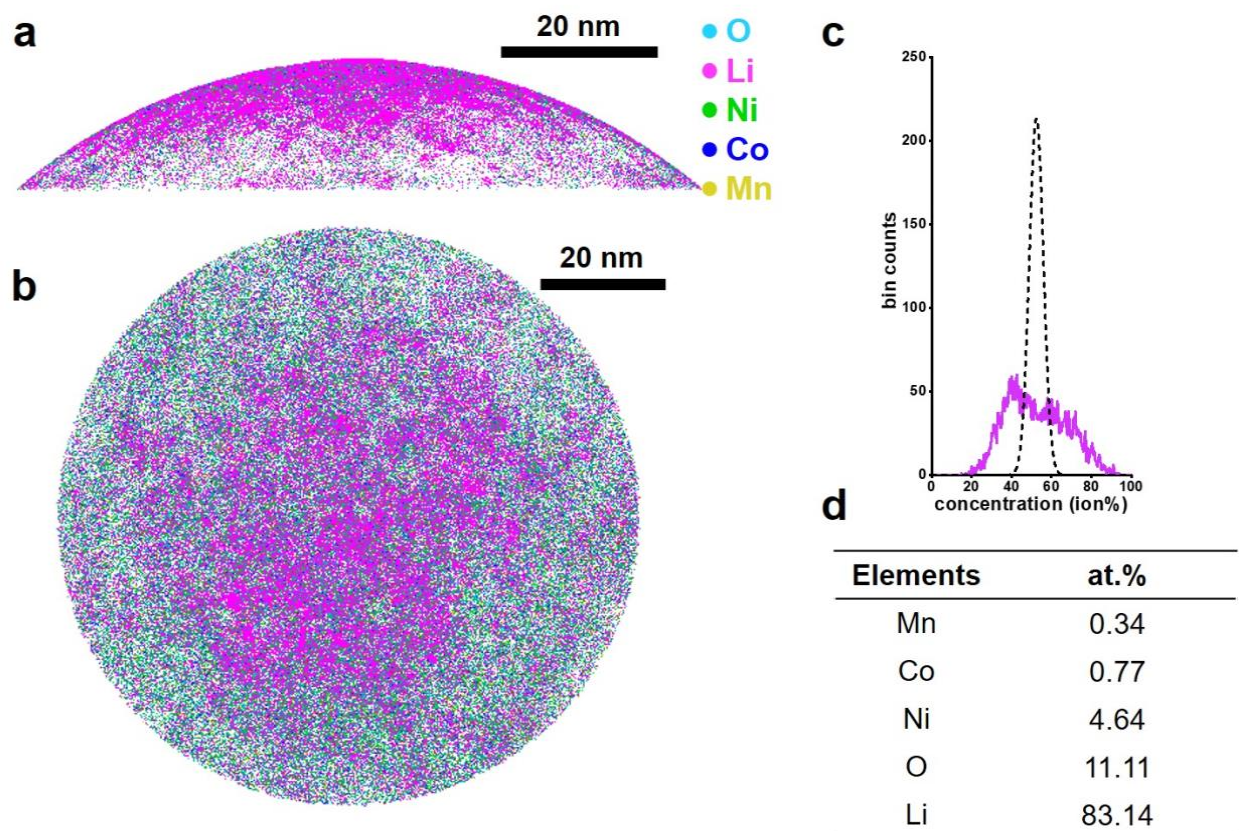

**Figure S11.** The air-transferred NMC811 sample measured at high laser energy (20pJ). (a) Side-view and (b) top-view 3D atom maps. (c) Measured (pink) and random (dashed line) binomial distribution of Li. (d) Overall compositional analysis. The measurement was manually stopped.

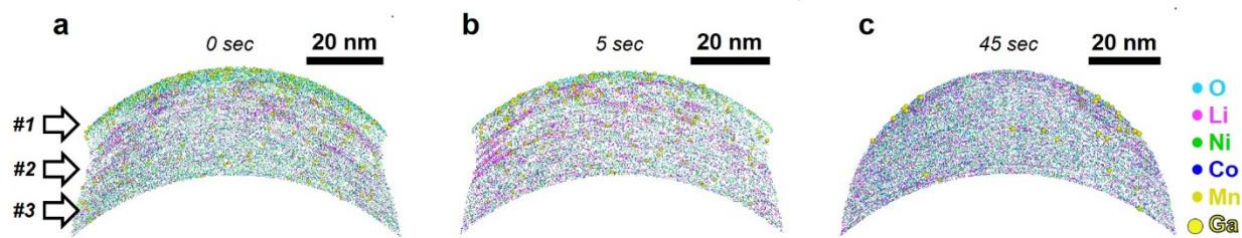

**Figure S12.** The three NMC811 specimens were prepared using Ga-FIB with different ion beam cleaning conditions to investigate the *(remaining)* Ga effect on the APT measurement. The low Ga ion beam at 5 kV and at 8 pA was used for (a) 0, (b) 5, and (c) 45 sec. All these specimens were transferred in air. Three distinguish regions were detected in the sample (a) and (b). #1 region contains high amount of Ga whereas #2 region showed some Li clusters region. But after 2M ions (region #3), the elements in the NMC811 specimens evaporated homogeneously without any indication of Li hotspots. In the case of the cleaned specimen (c), less Ga ions was detected. The measurement started with region #2 but it got improved to region #3 after <0.5 M ions. All measurement was stopped manually after 5M ions collection.

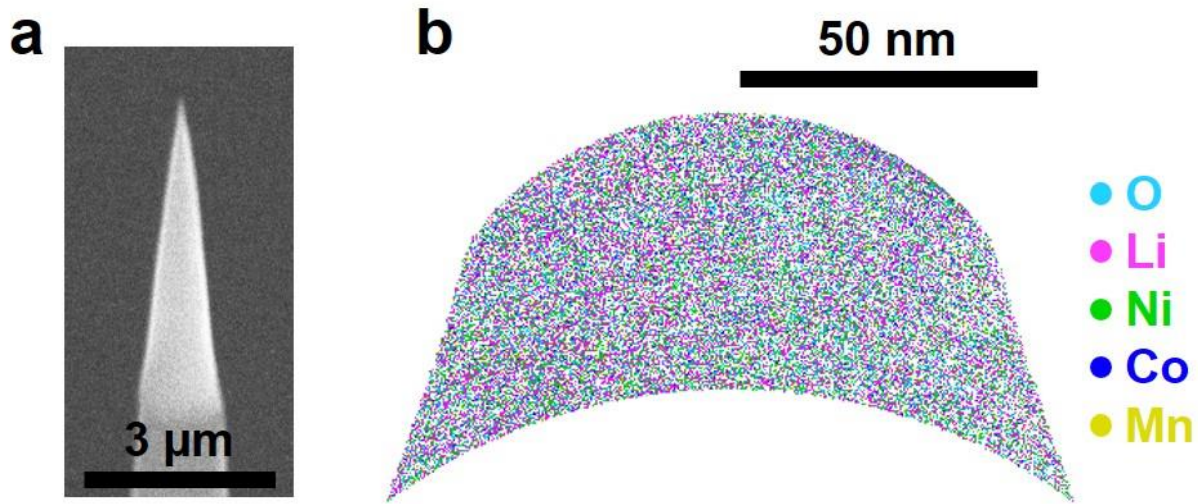

**Figure S13.** (a) NMC811 specimen fabricated using Xe-plasma FIB. (b) The corresponding APT result.

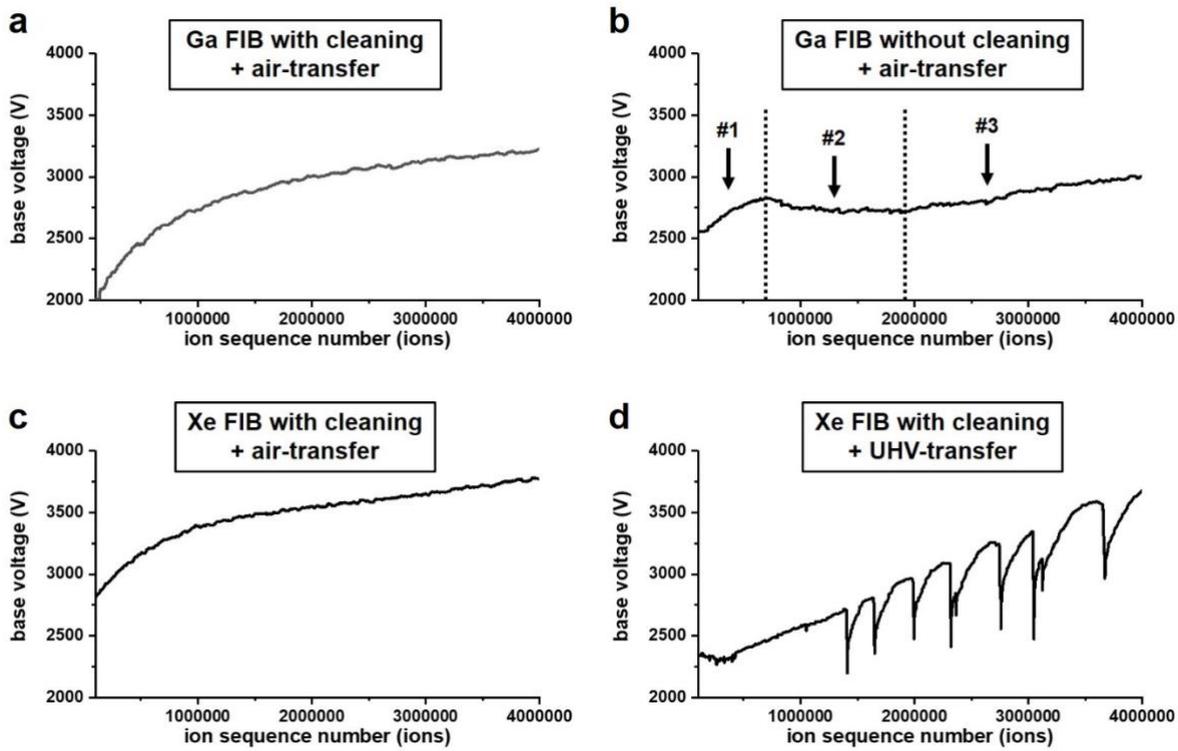

**Figure S14.** Voltage history curves of the NMC811 specimens fabricated using Ga- and Xe-FIBs: Ga-FIB (a) with and (b) without cleaning process. Xe-FIB (c) with air-transfer and (d) with UHV transfer. Note that there were many micro-fractures after UHV transfer and also note that there were three distinguished regions in the curve (b): Ga-concentrated region, Li-clustered region, and a *pristine* NMC811 region (see Figure S12a). All APT measurement was done at a base temperature of 60 K, a detection rate of 0.5 %, a pulse frequency of 125 kHz, and a laser energy of 5 pJ.

*Oxygen flux calculation*

The number of oxygen molecules (O₂) impinging on a sharpened specimen at different pressure can be calculated using the classical gas kinetic theory. The flux of O₂ (molecules m$^{-2}$ sec$^{-1}$) is given by:

$$\Phi = \frac{P}{\sqrt{2\pi M k_B T}}$$

, where $P$ is the gas pressure in Pa; here we assumed all remaining gases are O to yield an upper bound. Another assumption was made that all O gases will be adsorbed on the specimen surface (sticking factor = 1). $M$, $T$, and $k_B$ are the molecular weight of the gas in a.m.u., gas temperature in K, and the Boltzmann constant, respectively.

For the case of the (non-cryo) UHV transfer, the overall average pressure of the FIB chamber, intermediate chamber, and the suitcase, was ~10$^{-5}$ Pa. This value yields the flux of 14605 O₂ molecules m$^{-2}$ sec$^{-1}$. A typical APT specimen has a size of ϕ100 x 100 nm³. Assuming all O₂ molecules strike only onto a APT specimen, the O₂ adsorption rate will be 4.59 x 10$^{-10}$ molecules sec$^{-1}$ and it requires 69 days to achieve a monolayer coverage of O₂ on the specimen. Unlikely, when the ambient pressure value (e.g. the case of the air transferring) was inputted in the equation, the approximated time for monolayer development of O₂ on the battery specimen is 0.29 sec. This big difference in O₂ monolayer development time suggests that effect of reactive O₂ should be considered.

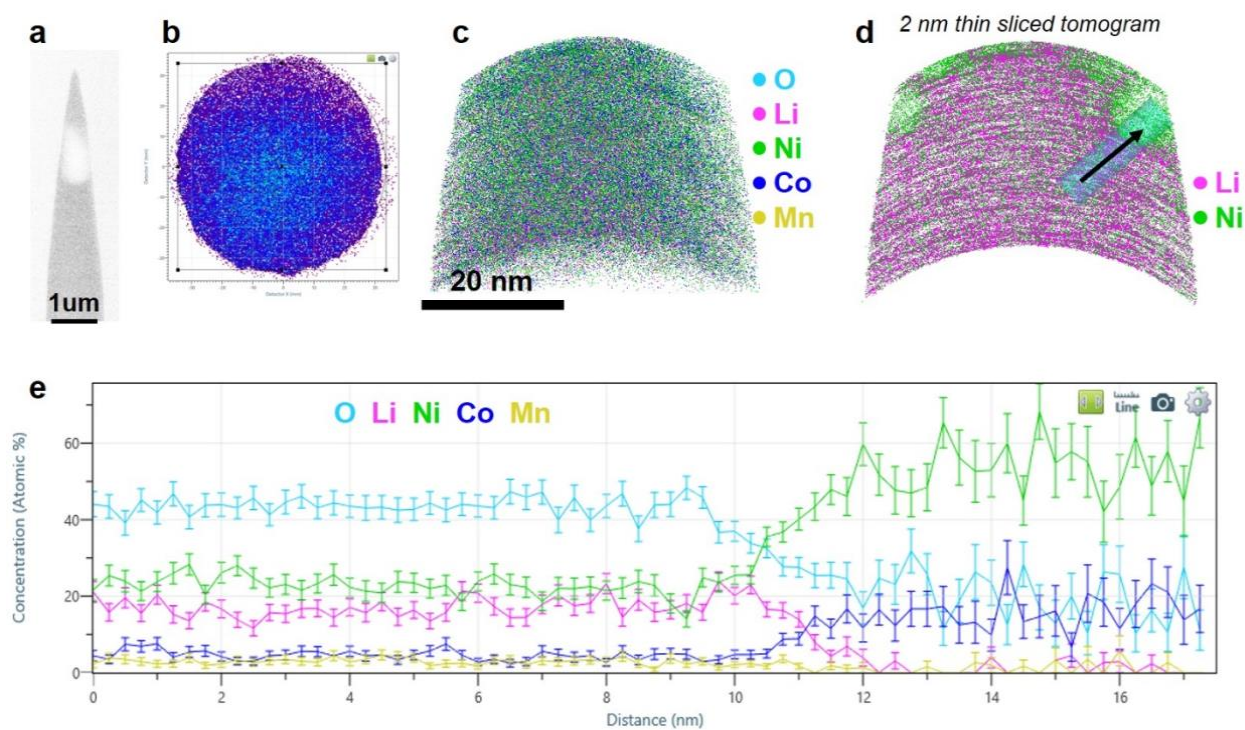

**Figure S15.** PVD-Ni coated NMC811 APT results with additional Ga milling after deposition: (a) APT specimen after PVD sputtering and Ga-FIB milling. (b) Overall density histogram. (c) 3D atom map of PVD-Ni. Cyan, pink, green, blue, and yellow dots represent reconstructed O, Li, Ni, Co, and Mn atoms, respectively. (d) 2 nm thin sliced along the x-axis. Note that Ni-rich region is detected that is PVD-Ni region. (e) 1D atomic compositional profile along cylindrical region of interest ($\phi$5 x 16 nm$^3$) in Figure S15d. Note that the Co mass-to-charge peak overlaps with the Ni isotope peak; therefore, it is not a Co element in PVD-deposited Ni.

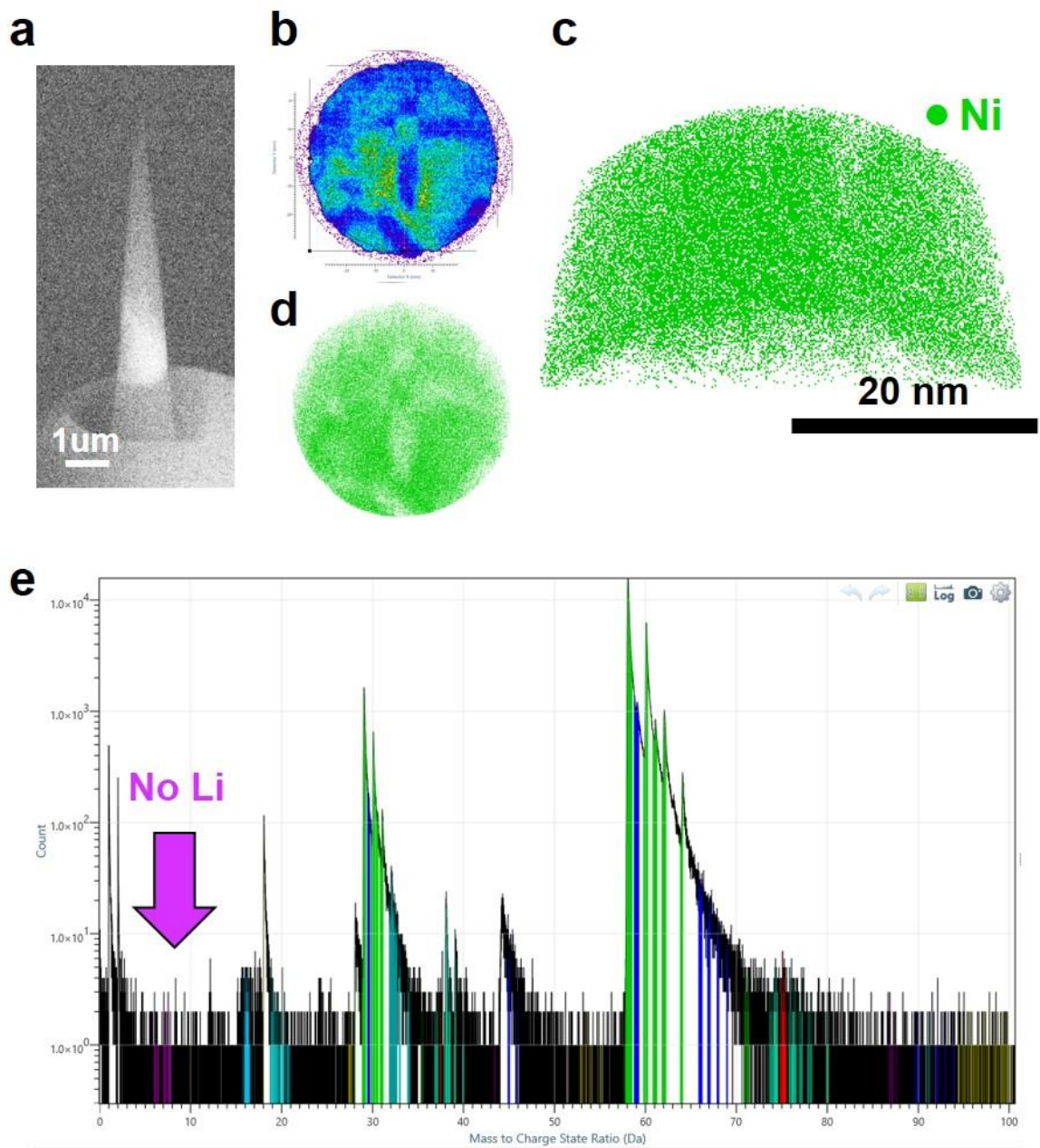

**Figure S16.** PVD-Ni coated NMC811 APT results: (a) APT specimen after PVD sputtering. (b) Overall density histogram. Note that pole figures are detected. (c) 3D atom map of PVD-Ni. Green dots represent Ni atoms. (d) 5 nm thin sliced along the measurement direction (z-axis). (e) Mass spectrum from acquired dataset. Note that no Li was detected indicating that Li did not electronically diffuse towards to the APT specimen apex.